\documentclass[aps,pra,twocolumn,a4paper,amssymb,amsmath,footinbib,showpacs]{revtex4}
\usepackage{multirow}
\usepackage{graphicx}
\usepackage{subfigure}
\usepackage{graphics}

\newcommand{\ket}[1]{| #1 \rangle}

\begin{document}

\title{Characterization of entanglement of more than two qubits \\
with Bell inequalities and global entanglement}
\author{Jochen \surname{Endrejat}}
\email[Electronic address:]{ jochen.endrejat@uni-bayreuth.de}
\author{Helmut \surname{B\"uttner}}
\affiliation{Theoretische Physik I, Universit{\"a}t Bayreuth, 
  D-95440 Bayreuth, Germany}
\date{\today}
\begin{abstract}
It is shown that the entanglement-structure of 3- and 4-qubit states
can be characterized by optimized operators of the Mermin-Klyshko type.
It is possible to discriminate between pure 2-qubit entanglements
and higher entanglements. A comparison with a global entanglement measure
and the i-concurrence is made.
\end{abstract}
\maketitle

\section{Introduction}

There seems to be no doubt in the literature that entanglement of
quantum mechanical states is one of the important ingredients in the
broad field of quantum information theory. Many protocols
in this area are based on entangled states, it is the basis
of quantum cryptography \cite{Ekert:91}, super dense coding \cite{Bennett:92},
teleportation \cite{Bennett:93} and other fields. 
Although it is an important ingredient
entanglement is even called puzzling, and especially genuine
multipartite entanglement for 3- and 4-qubits is a field of
active research, because only for two qubits a correct measure
for their entanglement is available. \\
In this paper we will give a new 
method of characterization of 3- and 4-qubit entanglement. \\
In the following we will show how different measures for quantifying
entanglement can be applied to a model spin system which can be used as
a basis for many different experimental setups. It is a one-dimensional
Heisenberg spin system with different arrangements for 3- and 4 spins.
We will compare a global entanglement measure \cite{MeyerWallach:02}
with the results of Bell
inequalities in the form proposed by Mermin and Klyshko 
\cite{Mermin:90:2,Klyshko:93,Belinski:93} which means
that we look for a measure with optimized polynomial spin operators.
It will be shown that the optimized polynomials measure the different degrees
of entanglement astonishing well.\\
The paper is structured as follows: 
In chapter II a special form of Bell inequalities
is described for 3- and 4-qubit systems.We also define
a new method how to handle these polynomial inequalities.
In chapter III we shortly discuss the global entanglement measure
and describe its relation to concurrences as well as to 3-qubit tangle measure.
In chapter IV these measures are applied to the general
Greenberger Horne Zeilinger (GHZ) state
for 3-qubits and the surprising result is that the optimized Mermin-Klyshko
operators are well suited to describe the entanglement as well as the global
entanglement over a wide range of parameters. In chapter V we now apply 
these different measures to the special form of a 3-qubit Heisenberg spin system
and again it turned out that for the pure eigenstates and for the superposition
of these eigenstates the optimized Bell operator as well as the global
entanglement measure are in excellent agreement describing the entanglement. 
In chapter VI we have a look at a 4-qubit system and compare it with the 3-qubit results.
A discussion follows in the concluding chapter VII.

\section{Spin-Polynomials for entanglement measure}

From Bell-Inequalities various polynomials are known for 
entanglement-classifications. We use here the Mermin-Klyshko
polynomials which were discussed by Yu et al. \cite{Yu:03} in their classififcation
scheme. We propose a special optimization procedure to give a quantitative
analyse of Heisenberg spin models.\\
We test our method for 3-qubit systems and then apply it to 4-qubits,
where it is shown that the resulting optimized
Mermin-Klyshko operators are able to quantitatively describe the total
entanglement measures. Furthermore, comparing it with the sum of the
2-qubit concurrences the difference gives a measure to the additional 3-
and 4-qubit entanglements. We note already here that the optimization
procedure yields many equivalent minima, therefore it is not possible to extract
directly a single well defined operator polynomial.

\subsection{3 qubits}

For 3-qubits the polynomials can be written as products
of spin operators \cite{Yu:03}:
\begin{align}
F_3  & = (AB'+ A'B)C + (AB-A'B')C'  \label{f3}\\
F'_3 & = (AB'+ A'B)C'- (AB-A'B')C \label{f3s} 
\end{align}
where the operators are written as sums of Pauli matrices.
\begin{equation*}
A^{(\prime)} = \vec{a}^{(\prime)} \cdot \vec{\sigma_A}, \quad
B^{(\prime)} = \vec{b}^{(\prime)} \cdot \vec{\sigma_B}, \quad
C^{(\prime)} = \vec{c}^{(\prime)} \cdot \vec{\sigma_C}
\end{equation*}
with $\vec{a}^{(\prime)}$, $\vec{b}^{(\prime)}$ and $\vec{c}^{(\prime)}$
normalised vectors and the Pauli matrices
$\vec{\sigma}_A$, $\vec{\sigma}_B$, $\vec{\sigma}_C$,
refering to the qubits $A$, $B$ and $C$,
with $\vec{\sigma_i}= (\sigma_i^x,\sigma_i^y,\sigma_i^z)$.\\
The classification for pure 3-qubit states is as follows.
(For details we refer to \cite{Yu:03}.)
We look for the maximum of the absolute values of the expectation
values of these polynomials and find for pure product states:
\begin{equation}\label{max1}
\text{max} \{|\langle F_3 \rangle_{\rho}|,|\langle F'_3 \rangle_{\rho}| \}\le 2
\end{equation}
The other inequalities found in \cite{Yu:03} can be written as
\begin{equation}\label{max2}
\langle F_3 \rangle^2_{\rho} + \langle F'_3 \rangle^2_{\rho} \le 2^3  
\end{equation}
if the state is 2-qubit entangled and
\begin{equation}\label{max3}
\langle F_3 \rangle^2_{\rho} + \langle F'_3 \rangle^2_{\rho} \le 2^4  
\end{equation}
if the state is 3-qubit entangled.\\
Our method consists in a numerical optimization 
of the components of the vectors 
$\vec{a}^{(\prime)}$, $\vec{b}^{(\prime)}$ and $\vec{c}^{(\prime)}$
of these polynomials.
We used the NAG library function e04ucc
\footnote{http://www.nag.co.uk} 
with randomly chosen initial conditions.\\
Since it was not clear from \cite{Yu:03} we applied two different methods.
In the first approach we look for the expectation value of $F_3$ and maximize it.
The polynomial $F'_3$ is calculated with these paramters and then
the sum $\langle F_3 \rangle^2 + \langle F'_3 \rangle^2$ is determined.\\
In the second approach we directly maximized the sum of the squares since all these
inequalities are sufficient but not necessary. By comparing the results of the
optimization with each other and with other measures of entanglement
we found that the first described method, the optimization 
of the $F_3 -$operator yields the best information.\\
Since we have not seen this numerical investigation even for the simplest states used
in the literature we cite the following results, obtained with the first
described method.\\
For the GHZ-state \cite{GHZ:89} $\bigl(\ket{000}+\ket{111}\bigr)/\sqrt{2}$:
\begin{alignat*}{2}
\text{max} \{|\langle F_3 \rangle_{GHZ}|\} &= 4.00 && \\
\Rightarrow |\langle F'_3 \rangle_{GHZ}|  &=0; &\quad
\langle F_3 \rangle^2_{GHZ} + \langle F'_3 \rangle^2_{GHZ} &= 16.00
\end{alignat*}
while for the so-called W-state \cite{Duer:00} 
$\bigl(\ket{001}+\ket{010}+\ket{100}\bigr)/\sqrt{3}$
we can list the following results which show that there 
seems to be a 3-party entanglement
although compared to GHZ it has not the maximum possible value.
\begin{alignat*}{2}
\text{max} \{|\langle F_3 \rangle_{W}|\} &= 3.05 && \\
\Rightarrow |\langle F'_3 \rangle_{W}|  &=0.05  &\quad
\langle F_3 \rangle^2_{W} + \langle F'_3 \rangle^2_{W} &= 9.305 
\end{alignat*}

\subsection{4 qubits}

The sufficient conditions for 4-qubits are a little more involved since we
have to introduce an additional spin polynomial for the $4^{\text{th}}$ qubit
but we can write altogether
\begin{align}
F_4 &= \frac{1}{2}(D+D')\otimes F_3 +\frac{1}{2}(D-D')\otimes F'_3 \\
F'_4 &= \frac{1}{2}(D+D')\otimes F'_3 +\frac{1}{2}(D'-D)\otimes F_3, 
\end{align}
where $F_3$ and $F^{\prime}_3$ are defined in (\ref{f3}) resp. (\ref{f3s})
and $D^{(\prime)}=\vec{d}^{(\prime)} \cdot \vec{\sigma_D}$ are the 
spin operators on the $4^{\text{th}}$ qubit $D$.
The classification scheme after Yu et al. \cite{Yu:03} is as follows. 
Product states fulfil the following inequality:
\begin{equation}
\text{max}\{|\langle F_4  \rangle _{\rho}|,|\langle F'_4  \rangle _{\rho}| \} \le 2 
\end{equation}
And for distinguishing different kinds of entanglement one can use 
the following scheme which gives sufficient but not necessary
classification.
\begin{itemize}
\item 2-qubit entanglement:
$\langle F_4 \rangle ^2_{\rho} + \langle F'_4 \rangle ^2_{\rho} \le 8$  
\item 3-qubit entanglement:
$\langle F_4 \rangle ^2_{\rho} + \langle F'_4 \rangle ^2_{\rho} \le 16$
\item 4-qubit entanglement:
$\langle F_4 \rangle ^2_{\rho} + \langle F'_4 \rangle ^2_{\rho} \le 32$,
\end{itemize}
where the description  \cite{Yu:03} is as follows.
4-qubit entanglement means a state with fully entangled 4-qubits,
3-qubit entanglement describes a product-state of one qubit with fully entangled
3-qubits, and 2-qubit entanglement can be a product of two 2-qubit entangled states
or a 2-qubit entangled state as product with two single qubits.\\
Before applying these inequalities to the spin-systems we discuss another
useful measure.

\section{i-concurrences and global entanglement}

The original measure $Q$ of a many qubit pure state $\ket{\psi}$
was introduced by Meyer and Wallach \cite{MeyerWallach:02}.
It was later shown by Brennen \cite{Brennen:03} that this kind
of global entanglement can be written as
\begin{equation}
Q(\ket{\psi})=2[1-1/n \sum_{k=1}^n Tr(\rho^2_k)], 
\end{equation}
with $\rho_k$, the density matrix reduced to a single qubit $k$.
It is interesting to note that there can be introduced
the so-called i-concurrence \cite{Rungta:01} which also is directly related to the reduced
density matrix $\rho_A$ of a subsystem $A$. This i-concurrence
measures the entanglement between two subsystems $A$ and $B$ and can be written as
\begin{equation}
IC_{A-B}=\sqrt{2[1-Tr(\rho^2_A)]}  
\end{equation}
In the following we use the notation $IC_{A-B} \equiv IC_A$.\\
We find as first result that the global entanglement
$Q$ is directly related to the sum of the squares of the i-concurrences
of the 1-qubit subsytems of a $N$ qubit state
\begin{equation}
Q=1/N \sum_{i=1}^N IC^2_i 
\end{equation}

\subsection{3 qubits}

For the special case of 3-qubits one can introduce the so-called tangle 
$\tau_{123}$ \cite{Coffman:00} which in a sense describes those contributions
to the i-concurrences which are not described by 2-qubit concurrences \cite{Hill:97,Wootters:98}
\begin{align}
IC^2_{1}=C^2_{12}+C^2_{13}+\tau_{123} \label{ic-conc}\\
IC^2_{2}=C^2_{12}+C^2_{23}+\tau_{123} \\
IC^2_{3}=C^2_{13}+C^2_{23}+\tau_{123} 
\end{align}
We can sum these relations up
\begin{align}
\sum_{i=1}^3 IC^2_{i} = 2(C^2_{12}+C^2_{13}+C^2_{23})+3\tau_{123} 
\end{align}
and introduce this into the global entanglement. It is nicely seen that for 3 qubits
this consists of the sum of squared 2-qubit concurrences plus the additional
tangle:
\begin{equation}\label{qconctau}
Q=\frac{2}{3} (C^2_{12}+C^2_{13}+C^2_{23}) + \tau_{123}   
\end{equation}
The total entanglement measure is the sum of different
entanglement contributions.

\subsection{4 qubits}

These nice results for 3-qubits cannot easily be extended to 4-qubits since
there is no equivalent definition of the corresponding higher tangle. But to give
an impression of the power of the description with a global measure
one can look for special qubit states
were there are effectively only 2-qubit concurrences.
\begin{equation}
\ket{\phi}=  \alpha_1 \ket{1000}+\alpha_2 \ket{0100}+
\alpha_3 \ket{0010}+ \alpha_4 \ket{0001} 
\end{equation}
One easily finds that the i-concurrences are sums of 2-qubits concurrences
and therefore the global entanglement can be written as
\begin{equation}\label{q-allgW}
Q = \frac{1}{2} \bigl( C_{12}^2 + C_{13}^2 + C_{14}^2 +
C_{23}^2 + C_{24}^2 + C_{34}^2 \bigr).
\end{equation}
Again this indicates a good total measure of entanglement 
by the value of $Q$.

\section{Application to generalized GHZ-state}

\begin{figure}
\includegraphics[width=5.5cm]{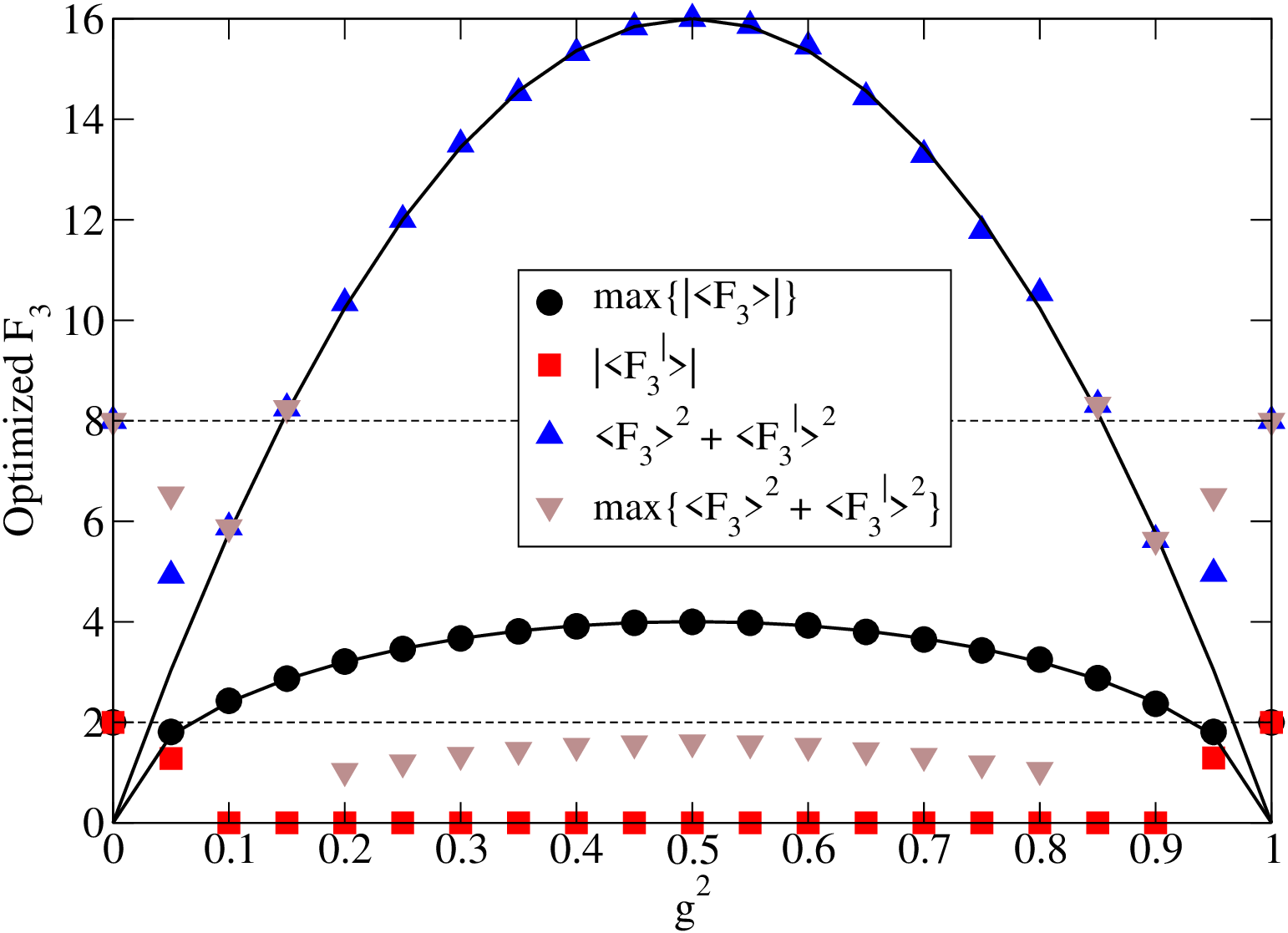}  
\caption{\label{bellvghz} Optimization of the expectation value of $F_3$ 
for the 3-qubit generalized GHZ-state as a function of $g^2$; the course
of $\text{max}\{|\langle F_3 \rangle|\}$ is fitted with $y = c_0 g \sqrt{1 - g^2}$,
with $c_0 = 8.01$; the course of $\langle F_3 \rangle^2 + \langle F'_3 \rangle^2$
with $y= c_1  g^2 (1-g^2)$ and $c_1=64.01$;
$\text{max}\{\langle F_3 \rangle^2 + \langle F'_3 \rangle^2 \}$ marks the
optimization of the squared inequalities. The dashed lines
mark the boundaries arising from the 
inequalities (\ref{max1}) resp. (\ref{max2}).
See text for more discussion.}
\end{figure}

One result of our investigations
is that the comparison of sufficient conditions from the Bell inequalities
and the global expression Q is an appropriate measure for the 
entanglement of 3- and 4-qubits.\\
As a first test we consider the generalised GHZ state for 3 qubits
written as
\begin{equation}
g \ket{000} + \sqrt{1-g^2} \ket{111},
\end{equation}
with $g \in [0,1]$.
It is well known that there are no 2-qubit concurrences so that the remaining
i-concurrences
\begin{equation}
IC_{1} = IC_{2} = IC_{3} = 2 g \sqrt{1 - g^2}
\end{equation}
mainly measure the tangle of the state
which is of course
parameter dependent and from our formula (\ref{qconctau}) it can be seen that
$Q$ just measures this tangle:
\begin{equation}\label{vghz-tangle}
Q = \tau_{123} = 4 g^2 (1-g^2)
\end{equation}
In fig. \ref{bellvghz} the results of the Bell optimization are plotted
as a function of $g^2$. $\text{max}\{|\langle F_3 \rangle|\}$ means the
optimization of the expectation value of $F_3$.
Then the resulting parameters are introduced in the expectation 
value of $F'_3$ (Notation: $|\langle F'_3 \rangle|$) 
and the squared inequalities (\ref{max2}) respectively (\ref{max3})
(Notation: $\langle F_3 \rangle^2 + \langle F'_3 \rangle^2$ ).
This procedure yields results that are nearly identical
to the parameter dependence of the i-concurrence and the tangle resp. $Q$
(cp. (\ref{vghz-tangle})) over a large parameter range.
This fact is underlined with the fit of the results of the inequalities by
the results of the i-concurrences resp. the tangle.
We therfore conclude that our maximization procedure yields
the correct information for the entanglement.
(Note however that the optimization procedure as with all 
search-algorithms may fail to find the appropriate maximum.)\\
Only for a small parameter range ($0 < g < 0.39$ and $0.92 < g < 1$) the inequalities are
not sufficient compared to the calculated tangle. These results show that 
at the boundaries of the parameter values the optimization of $F_3$ could
have problems but over a large range of the parameter values the optimized
$F_3$ gives a good measure of entanglement as the tangle itself although
we have no direct proof of the equivalence of these two meassures.
Quite remarkable is here the fact that the optimiztion of
$\langle F_3 \rangle^2 + \langle F'_3 \rangle^2$
(Notation: $\text{max}\{\langle F_3 \rangle^2 + \langle F'_3 \rangle^2 \}$)
yields for the parameter values $0.39 < g < 0.92$ no sufficient criterion for entanglement.\\
The points ($ g = 0.26$ and $g = 0.97$) where the inequality (\ref{max1}) is not violated agree 
with the results derived by Scarani and Gisin \cite{Scarani:01}.\\
This successful description of the generalized GHZ-states encourages us
to describe the entanglement of more complex systems.

\section{Pure 3-Qubit Heisenberg states}

As in reference \cite{Glaser:03} already discussed, Heisenberg
spin sytems are good models for various experimental realizations of
multi-qubit systems. Here we look for a special chain of 3
qubits where the interaction between qubit 1 and 2 and 2 and 3
are given by a certain interaction constant while between 1 and 3
we have doubled the interaction. Our main purpose is, however, to study
the anisotropy effect of this Hamiltonian given by
\begin{multline} 
H =\frac{J}{4}\Bigl(
\sigma_{1}^x \sigma_{2}^x + \sigma_{1}^y \sigma_{2}^y +
 \Delta \sigma_{1}^z \sigma_{2}^z +\\
\sigma_{2}^x \sigma_{3}^x + \sigma_{2}^y \sigma_{3}^y +
 \Delta \sigma_{2}^z \sigma_{3}^z \Bigr)  + \\
\frac{J}{2}\Bigl(
\sigma_{1}^x \sigma_{3}^x + \sigma_{1}^y \sigma_{3}^y +
 \Delta \sigma_{1}^z \sigma_{3}^z
 \Bigr), 
\end{multline}
with the anisotropy coefficient $\Delta$.
The eigensystem of this Hamiltonian is calculated
in the computational basis. The eigenvalues
and eigenstates are given in Table I. 

\begin{table}[b]
\caption{Eigensystem of the 3-qubit Hamiltonian}
\begin{ruledtabular}
\begin{tabular}{ll}
$E_1=\Delta J$ & $\ket{\psi_1} = \ket{111}$ \\
$E_2=\Delta J$ & $\ket{\psi_2} = \ket{000}$ \\
$E_3= - \frac{J}{2}$ &
$\ket{\psi_3}  =-\frac{1}{\sqrt{2}}\left( \ket{011} - \ket{110}\right)$ \\
$E_4= - \frac{J}{2}$ &
$\ket{\psi_4}  =-\frac{1}{\sqrt{2}}\left( \ket{001} - \ket{100}\right)$ \\
$E_5= - \frac{J}{4} (\eta + \Delta - 2)$ &
$\ket{\psi_5}  =\frac{\sqrt{\chi}}{2\sqrt{\eta}} 
\left( \ket{011} - \frac{4}{\chi} \ket{101} + \ket{110} \right)$ \\
$E_6= - \frac{J}{4} (\eta + \Delta - 2)$ &
$\ket{\psi_6}  =\frac{\sqrt{\chi}}{2\sqrt{\eta}} 
\left( \ket{100} - \frac{4}{\chi} \ket{010} + \ket{001} \right)$ \\
$E_7= \frac{J}{4} (\eta - \Delta + 2)$ &
$\ket{\psi_7}  =\frac{\sqrt{2}}{\sqrt{\eta} \sqrt{\chi}} \left(\ket{011} +
\frac{\chi}{2}  \ket{101} + \ket{110} \right) $ \\
$E_8= \frac{J}{4} (\eta - \Delta + 2)$ &
$\ket{\psi_8}  =\frac{\sqrt{2}}{\sqrt{\eta} \sqrt{\chi}} \left(\ket{100} +
\frac{\chi}{2}  \ket{010} + \ket{001} \right)$ \\ 
&\\
with: & \\
$\eta = \sqrt{12+\Delta(\Delta-4)}$ & $\;\text{and} \;\;\; \chi = \eta + \Delta -2 $ \\
\end{tabular}
\end{ruledtabular}
\end{table}

The eigenstates are of course partially degenerated because of the spin symmetry
of the system. This can be easily lifted by an applied field 
\begin{equation}
H' = H +  \frac{M}{2} \sum_{i=1}^3 \sigma_i^z,
\end{equation}
so that in the following we think of the different
eigenstates as pure states and discuss only the parameter dependent eigenstates
$\ket{\psi_5}$ to $\ket{\psi_8}$ because we are interested in the change of
entanglement with different anisotropy strengths.
Since the 2-qubit measures are known we will get more insight
into pure 3-qubit entanglement. As it was seen for the
generalized GHZ-state the parameter dependence of the states gives insight into the
efectiveness of different entanglement measures.

\subsection{The states $\ket{\psi_5}$, $\ket{\psi_6}$}

We start with the states $\ket{\psi_5}$ and $\ket{\psi_6}$
which yield the same results in measuring the entanglement.
The concurrences are calculated to be:
\begin{alignat}{2}
C_{12}=C_{23}&=\frac{2}{\eta} &\qquad
C_{13}&=\frac{4}{\eta^2 -(\Delta -2)\eta}.
\end{alignat}
In the limit $\Delta \to \infty$ the states have an easy form
\begin{align*} 
\ket{\psi_5} &\approx \frac{1}{\sqrt{2}}(\ket{01}_{13}+ \ket{10}_{13}) \otimes \ket{1}_2 \\
\ket{\psi_6} &\approx \frac{1}{\sqrt{2}}(\ket{01}_{13}+ \ket{10}_{13}) \otimes \ket{0}_2
\end{align*}
and the results for the concurrences are consistent with this form.
$C_{12}$ and $C_{23}$ are vanishing while $C_{13}$ increases to one.
The sum of the squared concurrences is needed for calculating $Q$
\begin{equation}
\sum C_{ij}^2 = 8 \Bigl(\frac{1}{\eta^2}+\frac{2}{(\eta^2-(\Delta-2)\eta)^2} \Bigr)
\end{equation}
To calculate the tangle $\tau_{123}$ one needs the i-concurrences
\begin{align}
IC_1 = IC_3 &= \frac{1}{\sqrt{2}}\sqrt{\smash[b]{1+\frac{4}{\eta^2}+\frac{\Delta -2}{\eta}}} \\
IC_2 &= \frac{2 \sqrt{2}}{\eta}
\end{align}
From these equations and the results for the concurrence it
follows with (\ref{ic-conc}) that $\tau_{123}=0$.
Therefore the total global entanglement $Q$ is given mainly by the
squares of the concurrences as it follows from equation (\ref{qconctau}),
and its parameter dependence is shown in fig. 2.

\subsection{The states $\ket{\psi_7}$, $\ket{\psi_8}$}

The states $\ket{\psi_7}$ and $\ket{\psi_8}$ have the same results for
entanglement measurement as well. The concurrences are calculated to
\begin{alignat}{2}
C_{12}= C_{23} & =\frac{2}{\eta} &\qquad
C_{13}& =\frac{4}{\eta^2 +(\Delta-2)\eta}.
\end{alignat}
In the limit $\Delta \to \infty$ the concurrences are vanishing,
the states have product form. For the special case $\Delta =1$,
$\ket{\psi_7}$ and $\ket{\psi_8}$ have the form of the W-state
\begin{align*}
\ket{\psi_7}= \frac{1}{\sqrt{3}}(\ket{011}+\ket{101}+\ket{110}) &= \ket{\widetilde{W}}\\
\ket{\psi_8}= \frac{1}{\sqrt{3}}(\ket{001}+\ket{010}+\ket{100}) &= \ket{W}  
\end{align*}
and the three concurrences are identical. The sum of the squared concurrences
is needed for further calculations
\begin{equation}
\sum C_{ij}^2 = 8 \Bigl(\frac{1}{\eta^2}+\frac{2}{(\eta^2+(\Delta-2)\eta)^2} \Bigr) 
\end{equation}
From the i-concurrences 
\begin{align}
IC_{1-23} = IC_{3-12} &=\frac{1}{\sqrt{2}} 
\sqrt{\smash[b]{1 + \frac{4}{\eta^2}+\frac{2-\Delta}{\eta}}} \\
IC_{2-13} &= \frac{2 \sqrt{2}}{\eta}
\end{align}
follows with (\ref{ic-conc}) that $\tau_{123}=0$. 
Again there is no genuine 3-qubit entanglement as the $\tau -$measure
indicates. In addition the $Q-$measure is calculated
(shown in fig. \ref{ps5ps7-plot}). Again, the global entanglement
$Q$ is only a function of the sum of the squared concurrences, which
follows directly from equation (\ref{qconctau}).\\
In the next section we compare these results with the optimized
inequalities.

\subsection{Comparison with Bell optimization}

\begin{figure}
\includegraphics[width=5.5cm]{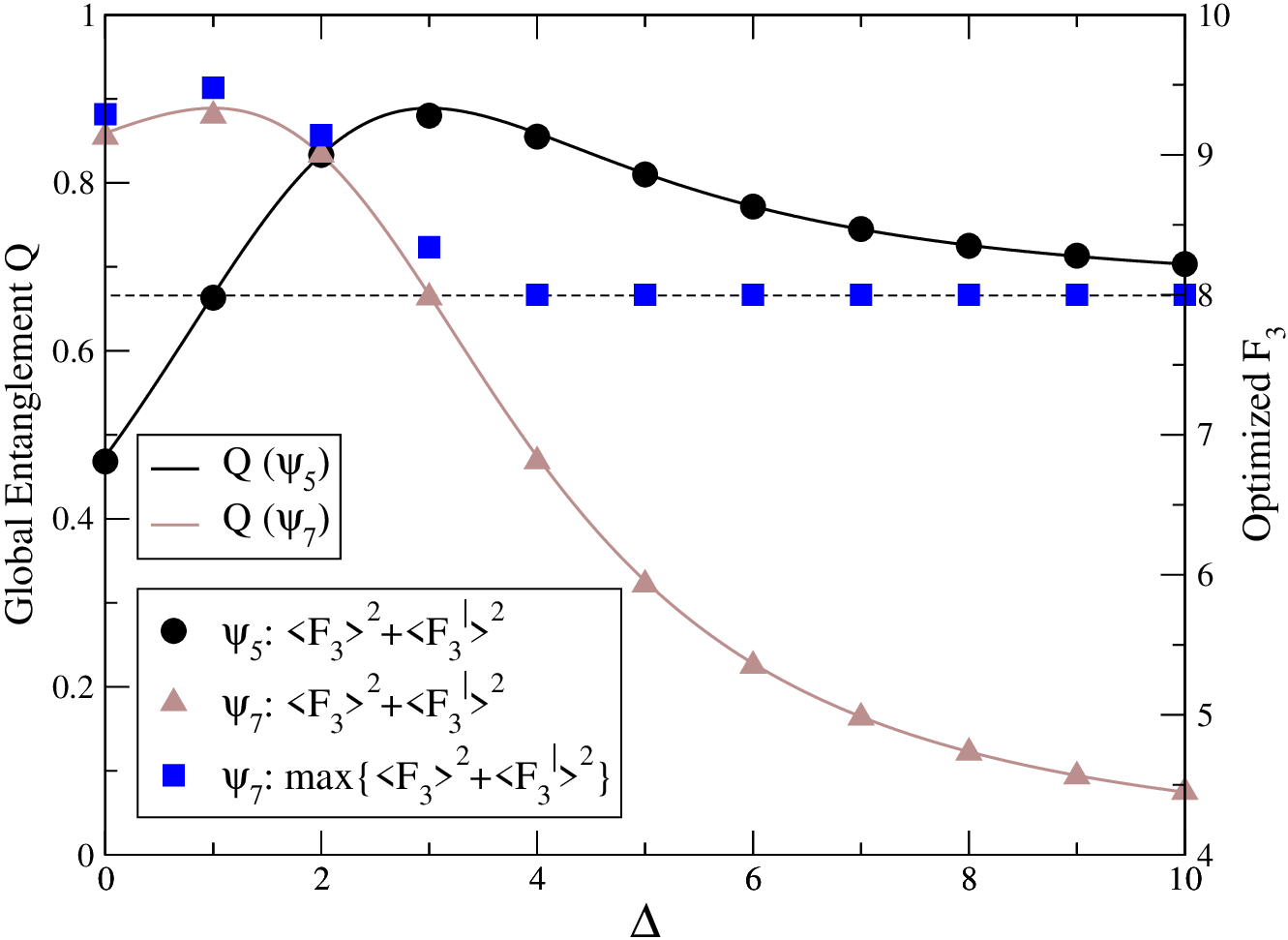}  
\caption{\label{ps5ps7-plot}Comparison of global entanglement $Q$ and the results of the
$F_3$ optimization for the states $\ket{\psi_5}$ and $\ket{\psi_7}$; 
additionally the optimization of $\langle F_3 \rangle^2 + \langle F'_3 \rangle^2$
for the state $\ket{\psi_7}$ is plotted.}
\end{figure}

In fig. \ref{ps5ps7-plot} besides the global entanglement $Q$ we plot the results
of the Bell optimization as a function of the anisotropy
parameter $\Delta$. The left y-axis shows $Q$, the right one the results for
the squared inequalities. For both pairs of states 
$\ket{\psi_5}$, $\ket{\psi_6}$ and $\ket{\psi_7}$,  $\ket{\psi_8}$
the inequality (\ref{max1}) is violated. This is a sufficient condition for
entanglement. With the squared inequalities (\ref{max2}) and (\ref{max3})
we can distinguish 2-qubit and 3-qubit entanglement. The states
$\ket{\psi_5}$ and $\ket{\psi_6}$ are for $\Delta > 1.03$
3-qubit entangled. The states $\ket{\psi_7}$ and $\ket{\psi_8}$
show 3-qubit entanglement in the range $0 \le \Delta \le 2.97$.
The points that mark the transition between 2-qubit and 3-qubit entanglement were
determined by $\langle F_3 \rangle^2_{\rho} + \langle F'_3 \rangle^2_{\rho}=8.01$.\\
These results are in accordance with the 3-qubit classification after 
D\"ur et al. \cite{Duer:00}. There is the so-called W-class of states
which are 3-qubit entangled and the tangle is 0.\\
But if we compare the course of the optimized $F_3$ with
the course of the global entanglement $Q$ we find - up to an scaling factor -
exact analogy. This indicates that an apparent 3-qubit entanglement
is due to to the fact that the sum of the squared concurrences is larger than 1,
the global entanglement $Q$ larger than $2/3$, cp. (\ref{qconctau}).
In the discussion of these parameter dependent states an
interesting result is that although both states have no finite
tangle they differ in the aspect of the strength of the
2-qubit entanglement as measured by the sum of the squares of the concurrence.
As soon as this sum is larger than 1 then there seems
to be a kind of effective 3-qubit entanglement which is
measured by the optimized $F_3$ operator.
This means, that besides the pure 3-qubit tangle one has to consider
the ``strength'' of the 2-qubit total concurrence, which might
effectively describe some indirect 3-qubit entanglement.
(But not a genuine one as measured by the tangle.)
This means in our interpretation that 3-qubit states with tangle equals to 0 are
only 2-qubit entangled.\\
With our results we conclude that also the W-state
is only 2-qubit entangled, because the sum of the
squarred concurrences is equal to $4/9$, cp. \cite{Duer:00}.\\
In fig. \ref{ps5ps7-plot} we plotted additionally the optimization
of the squared inequalities for the state $\ket{\psi_7}$. 
For $\Delta \ge 4$ the optimization yields
$\text{max}\{\langle F_3 \rangle^2 + \langle F'_3 \rangle^2\} =8.00$
indicatig 2-qubit entanglement. 
As one can see, this method yields the same sufficient conditions due to entanglement
classification, but less information due to entanglement measurement.\\  
In the following we will discuss the superposition of two states
in order to create a state with a finite tangle and to find at the same time the optimized
$F_3$ operator.

\subsection{Superposition of $\ket{\psi_7}$ and $\ket{\psi_8}$}

\begin{figure}[!b]
\includegraphics[width=6cm]{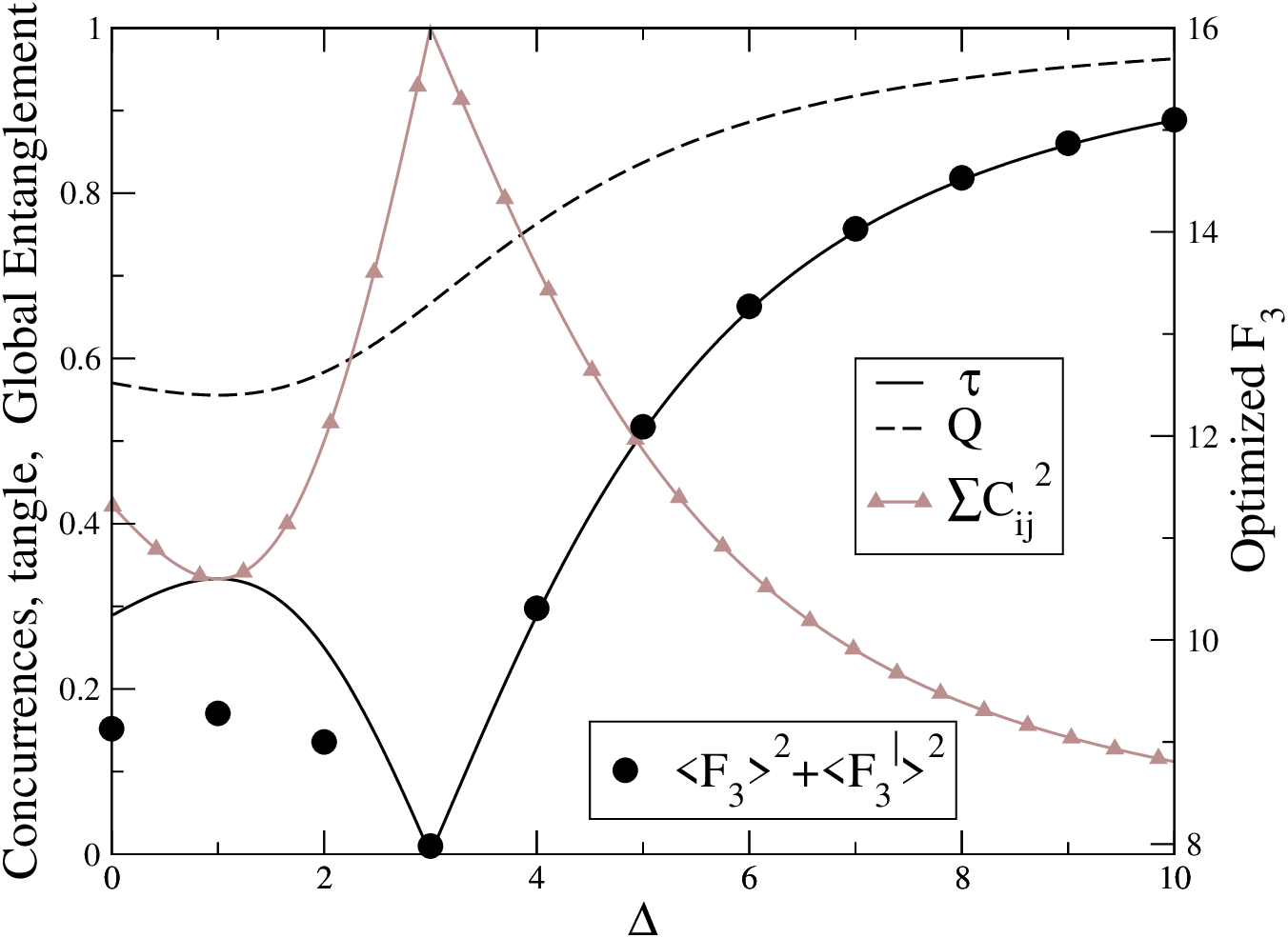}
\caption{\label{erg-ps78} 
Measurement of the superposition of $\ket{\psi_7}$ and $\ket{\psi_8}$;
left y-axis: global entanglement, sum of squared concurrences and tangle;
right y-axis: optimized $F_3$ operator.}
\end{figure}

In this part we will discuss the superposition of the 
degenerated states $\ket{\psi_7}$ and $\ket{\psi_8}$:
\begin{equation*}
\frac{1}{\sqrt{2}} \bigl( \ket{\psi_7} + \ket{\psi_8} \bigr)  
\end{equation*}
In the limit $\Delta \to  \infty$ we get a highly entangled state
\begin{equation}
\frac{1}{\sqrt{2}}( \ket{\psi_7}+\ket{\psi_8})
\approx \frac{1}{\sqrt{2}}(\ket{010} + \ket{101})
\end{equation}
The concurrences, the i-concurrences and the tangle 
can be calculated exactly.
The expressions for the concurrence and the tangle
are quite long and because of simplicity we will discuss
them only graphically.\\
The calculation of $Q$ gives the following result:
\begin{equation}
Q =\frac{12+5\eta^2+(\Delta-2)\eta}{6\eta^2}
\end{equation}
In fig. \ref{erg-ps78} we have plotted the sum of the squared concurrences, 
the global entanglement $Q$ and the tangle $\tau$ as a function of $\Delta$
on the left y-axis. The right y-axis shows the results for the 
optimization of $F_3$.\\
This superposition of two degenerated states has finite tangle and the sum
of the squared concurrences is never larger than 1. From fig. \ref{erg-ps78}
it is interesting to note that at least above $\Delta =3$ there is a perfect
agreement between the tangle and the optimized $F_3$ operator. 
Below $\Delta =3$ a tendency of the 3-tangle is reproduced.
To sum it up it can be said that the $F_3$ structure shows tangle like results
and therefore measures genuine 3-qubit entanglement.\\
Altogether one should also note that now the global entanglement measure 
$Q$ sums up all different kinds of entanglement, the 2-qubit entanglement
as measured by the sum of the squared concurrences and the 3-qubit entanglement
as measured by the tangle, cp. (\ref{qconctau}).\\
It is remarkable in this respect that the $F_3$ operator in its optimized form
has an interesting structure. E.g. for large $\Delta$ where the state
is mainly a GHZ type state, the operator mainly consists of linear combinations
of $\sigma^x$ and $\sigma^y$. There is no contribution from the $\sigma^z$ components,
but it is important to note that although the $\sigma^x$ contributions are
small, they are important in the description of the actual entanglement. This is
seen by the fact that if one decreases $\Delta$ the entanglement decreases and this
is seen by the fact that now the $\sigma^x$ contributions get much stronger although
the $\sigma^z$ components still are negligible. Only for smaller $\Delta$, namely
in the region of $\Delta =3$ where the tangle goes to 0, one clearly sees that our optimized
operator has now quite large contributions from $\sigma^z$. Looking into the calculation
of the tangle one can conclude from this that a finite contribution in $F_3$ coming
from the $\sigma^z$ operators may indicate a small genuine 3-qubit entanglement.\\
This gives us sufficient confidence to discuss now 4-qubit systems where an
explicit measure of 3- and 4-qubit entanglement is not known.
And it turns out that the $F_4$ optimization
will yield additional information. In order to compare with the 3-qubit
results we use this time a special isotropic system ($\Delta =1$) and couple
the $4^{\text{th}}$ spin with a different coupling constant $J_s$.

\section{Pure 4-Qubit Heisenberg states}

The Hamiltonian of our 4-qubit system can be written as
\begin{equation}
H = \frac{J}{4} (\vec{\sigma_1}\vec{\sigma_2} +\vec{\sigma_2}\vec{\sigma_3})
+ \frac{J_s}{4} (\vec{\sigma_2}\vec{\sigma_4}),
\end{equation}
with the product $\vec{\sigma_i}\vec{\sigma_j}= \sigma^x_i \sigma^x_j +
\sigma^y_i \sigma^y_j + \sigma^z_i \sigma^z_j$.
The coupling $J_s$ between spin 2 and 4 attaches the $4^{\text{th}}$ spin
to the 3 qubits interacting homogeneously. One can easily determine the
eigenenergies and states of the system. We find out that two states are
of special interest and call them $\ket{\Phi_1}$ and $\ket{\Phi_2}$.
They are energetically degenerated and belong to a spin triplet. The 
abbreviations we use in the following parts, are given in table II.

\begin{table}[!b]
\caption{Abbreviations for the 4-qubit model}
\begin{ruledtabular}
\begin{tabular}{ll}
$\delta$  $= \sqrt{9 J^2 -4 J J_s + 4 J_s^2} $ &
$\mu_2 $  $= \frac{1}{\sqrt{3+\frac{9J-2J_s}{\delta}}}$\\
$a_1$  $=\frac{1}{2\sqrt{2}\mu_2} $ &
$b_1$  $=\frac{J_s -J}{\sqrt{2}\mu_2 (3J+\delta)}$\\
$c_1$  $=\frac{\mu_2(3J+2J_s+\delta)}{2\sqrt{18J^2-8JJ_s+8J_s^2}}$ & 
$a_2$  $\sqrt{4+\frac{(-J+2J_s+\delta)^2}{2J^2}}$
\end{tabular}
\end{ruledtabular}
\end{table}

\subsection{The state $\ket{\Phi_1}$}

\begin{figure*}
\includegraphics[width=12cm]{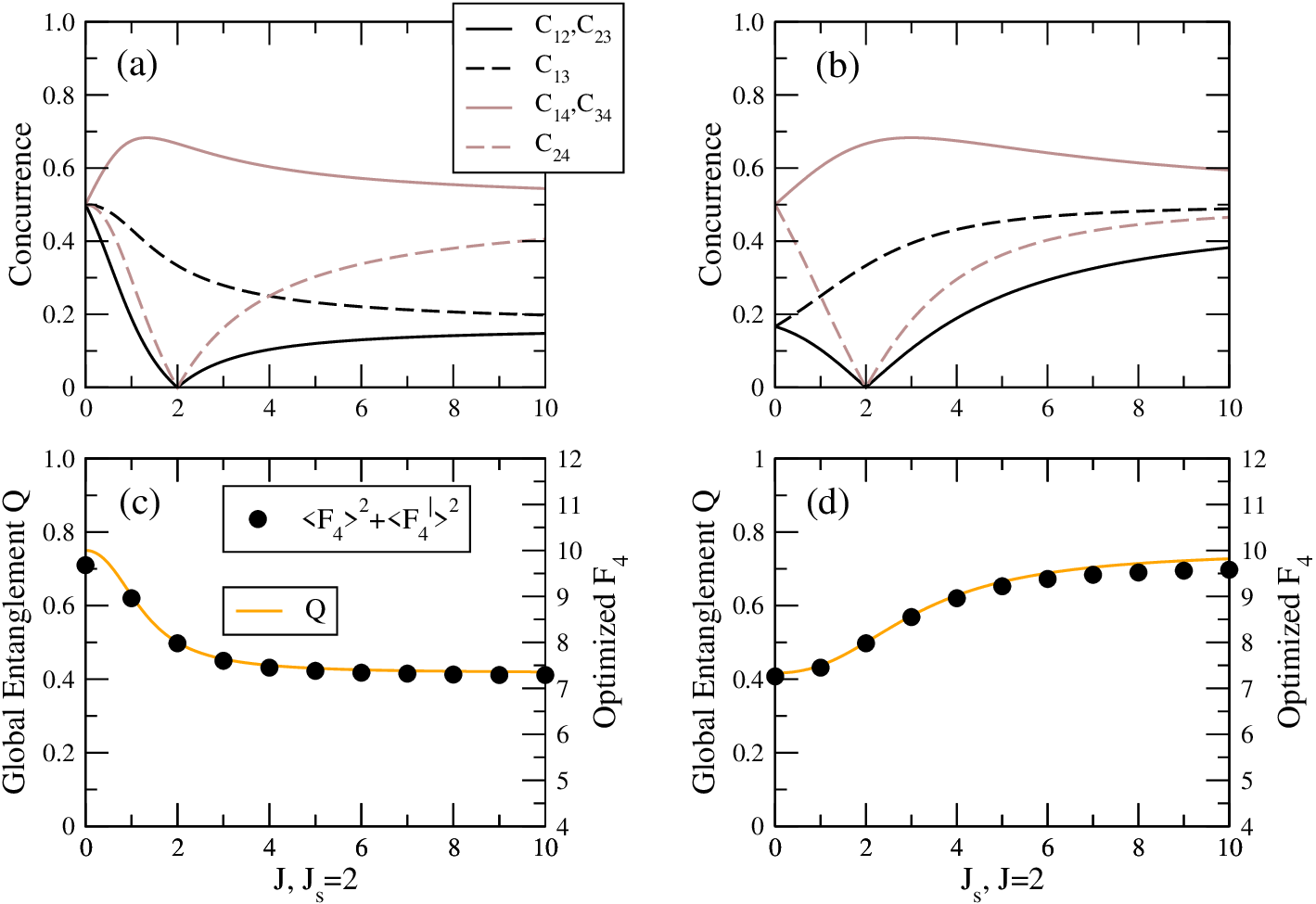}
\caption{\label{plot1416} Concurrences, global entanglement
and $F_4$ optimization for the state $\ket{\Phi_1}$;
(a): $C_{ij} (J, J_s=2)$; (b): $C_{ij} (J_s, J=2)$;
(c): left y-axis: $Q$, right y-axis: $F_4$ optimization, as a function of $J$, $J_s=2$;
(d): left y-axis: $Q$, right y-axis: $F_4$ optimization, as a function of $J_s$, $J=2$.}
\end{figure*}

The state $\ket{\Phi_1}$ is of generalized W form and written as
\begin{equation}
\ket{\Phi_1} = a_1\ket{1110} +b_1\ket{1011}+c_1\ket{0111}-c_1\ket{1101}.  
\end{equation}
Because of this structure it is clear from equation (\ref{q-allgW})
that the entanglement of this state is completely described by 2-qubit concurrences.
These concurrences have been calculated in the following form:
\begin{align}
C_{12} = C_{23} &= \frac{1}{2\sqrt{2}}
\sqrt{1+\frac{8J(J-J_s)}{\delta^2}+\frac{-5J+2J_s}{\delta}} \\
C_{13} &= \frac{1}{2\sqrt{2}}
\sqrt{1-\frac{4J^2}{\delta^2}+\frac{-J+2J_s}{\delta}} \\
C_{14} = C_{34} &= \frac{3J+2J_s +\delta}{4 \delta} \\
C_{24}&= \frac{(J-J_s)(3+\frac{9J-2J_s}{\delta})}{6J+2\delta}
\end{align}
Their dependences on the parameters $J$ and $J_s$ are given in fig. \ref{plot1416}a,b.
With these results it is easy to sum the squares with the result given by 
\begin{equation}
\sum C_{ij}^2 = \frac{1}{4}\Bigl(5-\frac{12J^2}{\delta^2}+
\frac{-J+2J_s}{\delta}\Bigr).
\end{equation}
and the calculation of the global entanglement (cp. (\ref{q-allgW})) yields
(see fig. \ref{plot1416}c,d)
\begin{equation}
Q=\frac{1}{8}\Bigl(5-\frac{12J^2}{\delta^2}+
\frac{-J+2J_s}{\delta}\Bigr).
\end{equation}
From these figures various conclusions can be drawn. First of all, there are special points
in the parameter space where certain concurrences are 0, especially for $J=J_s=2$. Form this one
can conclude that the qubit 2 can be separated in the state which indeed is true.
\begin{equation*}
\ket{\Phi_1} = -\frac{1}{\sqrt{6}}\Bigl( \ket{011}_{134}+\ket{101}_{134}-
2\ket{110}_{134} \Bigr) \otimes \ket{1}_2
\end{equation*}
Furthermore,
it can be seen that a total entanglement decreases as a function of $J$
and increases monotonically as a function of $J_s$ which means that by coupling these
3 qubits to a  $4^{\text{th}}$ one in this special state we can increase the total
entanglement which can be helpful in cetain experimental situations.
But most interestingly
when we compare these results with the Bell inequality result, we find that the optimized
Mermin-Klyshko polynomial operator $F_4$ is up to a scaling factor the same 
function as the total global entanglement. 
But differently from this factor in the case of $F_4$ one can extract the information 
that for $J_s=2$ and $J \lesssim 1.94$ as well as for $J=2$ and $J_s \gtrsim 2.06$
the sum of the quadratic concurrences lies above 1 which equals to the fact that
we have $\langle F_4 \rangle^2 + \langle F'_4 \rangle^2 > 8$ which indicates
an effective 3-qubit entanglement as described above due to the large
sum of the squared concurrences.\\
Another interesting aspect can be observed when the interaction 
constants are 0. When looking at the fig. \ref{plot1416} a,b 
all the 2-qubit concurrences are unequal to 0 for $J_s=2, J=0$
and $J=2, J_s=0$. At these values the entanglement is due to symmetry
effects and not arising from interaction. 

\subsection{The state $\ket{\Phi_2}$}

\begin{figure*}
\subfigure[]{\includegraphics[width=6cm]{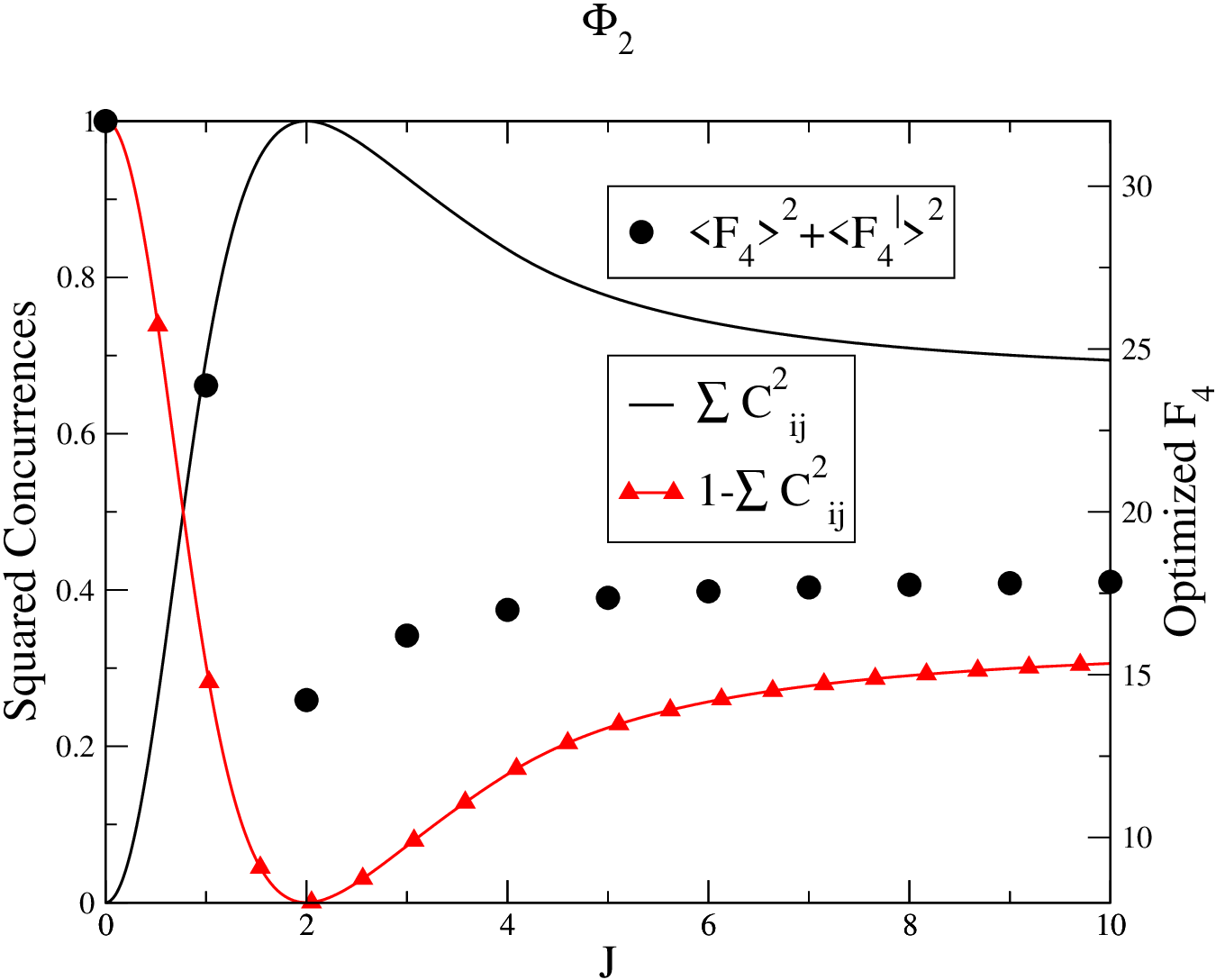}}  
\hfil
\subfigure[]{\includegraphics[width=6cm]{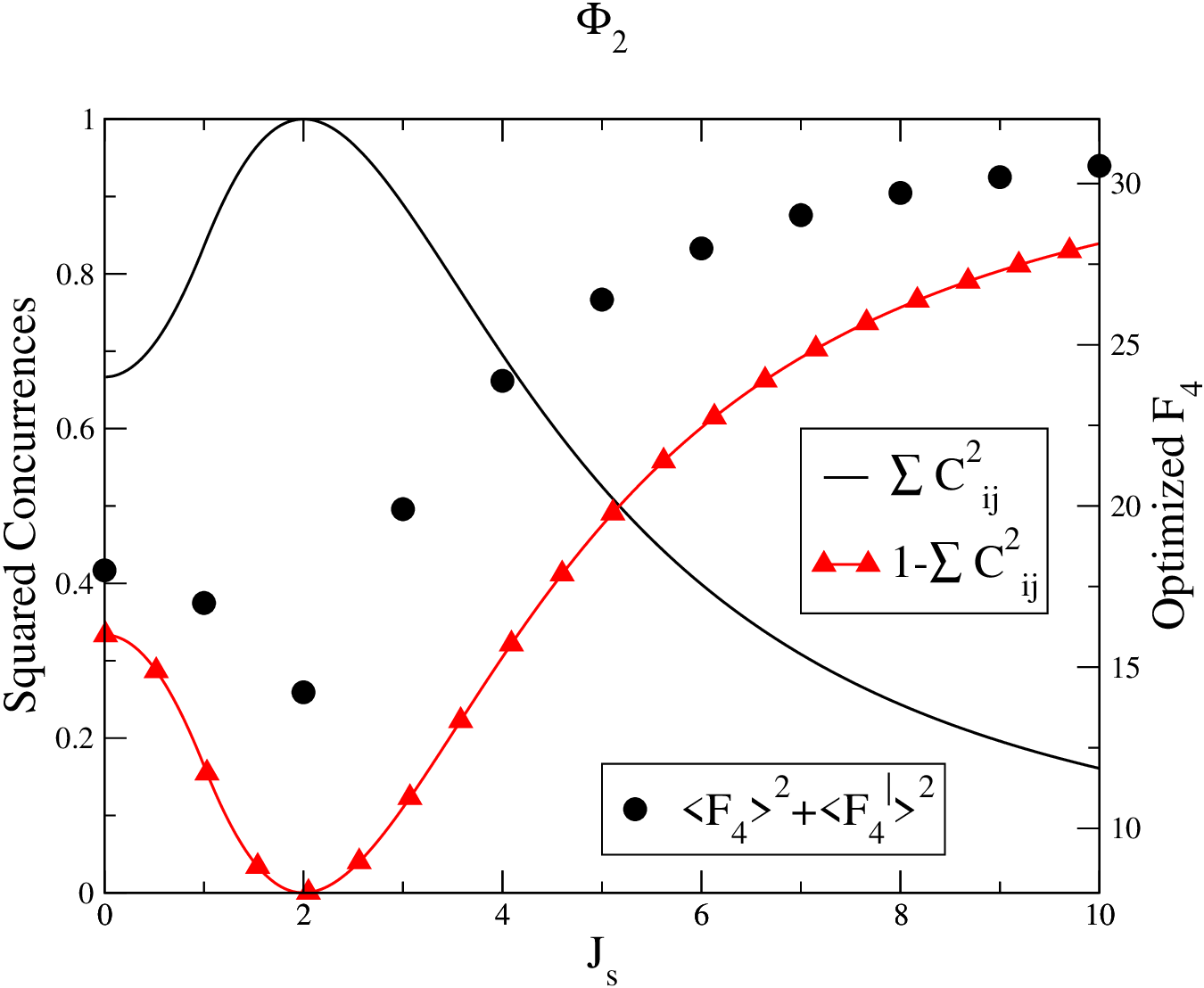}}  
\caption{\label{phi15-erg}Comparison of the sum of the squared concurrences and
the $F_4$ optimization for the state $\ket{\Phi_2}$.}
\end{figure*}

\begin{table*}
\caption{\label{bellparameter}Parameters out of the $F_4$ optimization for the state $\ket{\Phi_2}$;}
\begin{ruledtabular}
\begin{tabular}{ccc||ccc||ccc||ccc}
\multicolumn{6}{c||}{$J_s=2, J=0$}&\multicolumn{6}{c}{$J_s=0, J=2$}\\ \hline
\multirow{3}{5mm}{$A$} &$\sigma_A^x$& 7.33587e-01  &\multirow{3}{5mm}{$A'$} &$\sigma_{A'}^x$  &-6.79595e-01 &\multirow{3}{5mm}{$A$}&$\sigma_A^x$& 7.52415e-01&\multirow{3}{5mm}{$A'$}&$\sigma_{A'}^x$ &3.37233e-01\\
                       &$\sigma_A^y$& 6.79595e-01  &			    &$\sigma_{A'}^y$  &7.33587e-01  &                      &$\sigma_A^y$& 5.16326e-01&		       &$\sigma_{A'}^y$  &2.31417e-01 \\
                       &$\sigma_A^z$& 2.11433e-07  &			    &$\sigma_{A'}^z$  &2.93477e-07  &  	        &$\sigma_A^z$&-4.08998e-01&			 &$\sigma_{A'}^z$&9.12535e-01     \\ \hline
\multirow{3}{5mm}{$B$} &$\sigma_B^x$& 6.72877e-01  &\multirow{3}{5mm}{$B'$} &$\sigma_{B'}^x$  &7.39754e-01  &\multirow{3}{5mm}{$B$}&$\sigma_B^x$& 8.07789e-01&\multirow{3}{5mm}{$B'$} &$\sigma_{B'}^x$  &1.65322e-01 \\ 
                       &$\sigma_B^y$& 7.39754e-01  &			    &$\sigma_{B'}^y$  &-6.72877e-01 &		           &$\sigma_B^y$& 5.54324e-01&				&$\sigma_{B'}^y$  &1.13448e-01 \\
                       &$\sigma_B^z$& 1.93281e-07  &			    &$\sigma_{B'}^z$  &2.55498e-07  &		           &$\sigma_B^z$&-2.00504e-01&				&$\sigma_{B'}^z$  &9.79693e-01 \\ \hline 
\multirow{3}{5mm}{$C$} &$\sigma_C^x$& 6.28202e-01  &\multirow{3}{5mm}{$C'$} &$\sigma_{C'}^x$  &7.78050e-01  &\multirow{3}{5mm}{$C$}&$\sigma_C^x$& 7.43104e-01&\multirow{3}{5mm}{$C'$} &$\sigma_{C'}^x$  & 3.57282e-01 \\ 
                       &$\sigma_C^y$&-7.78050e-01  &			    &$\sigma_{C'}^y$  &6.28202e-01  &		           &$\sigma_C^y$& 5.09936e-01&				&$\sigma_{C'}^y$  & 2.45176e-01\\ 
                       &$\sigma_C^z$&-1.48041e-07  &			    &$\sigma_{C'}^z$  &-1.92036e-07 &		           &$\sigma_C^z$&-4.33315e-01&				&$\sigma_{C'}^z$  & 9.01242e-01 \\ \hline
\multirow{3}{5mm}{$D$} &$\sigma_D^x$& 1.90480e-01  &\multirow{3}{5mm}{$D'$} &$\sigma_{D'}^x$  &9.81691e-01  &\multirow{3}{5mm}{$D$}&$\sigma_D^x$& 2.32613e-01&\multirow{3}{5mm}{$D'$} &$\sigma_{D'}^x$  &7.91041e-01 \\ 
                       &$\sigma_D^y$& 9.81691e-01  &			    &$\sigma_{D'}^y$  &-1.90481e-01 &		           &$\sigma_D^y$& 1.59625e-01&				&$\sigma_{D'}^y$  &5.42831e-01 \\ 
                       &$\sigma_D^z$&-1.79827e-07  &			    &$\sigma_{D'}^z$  &3.21086e-07  &		           &$\sigma_D^z$& 9.59381e-01&				&$\sigma_{D'}^z$  &-2.82115e-01 \\ 
\end{tabular}
\end{ruledtabular}
\end{table*}

Even more interesting are the entanglement characteristics fot the state $\ket{\Phi_2}$.
We will apply our reasoning also to this state although one can only give partially quantitative
answers. First of all we note that the general form of the state given as
\begin{multline}
\ket{\Phi_2} = -\frac{1}{a_2}\ket{0011}+\frac{1}{a_2}\ket{0110}\\
-\frac{1}{a_2}\ket{1001}+\frac{1}{a_2}\ket{1100}\\  
-\frac{J a_2}{2 \delta}\ket{0101}+\frac{J a_2}{2 \delta}\ket{1010}  
\end{multline}
is quite complicated but it reduces to a GHZ state in the two limits 
$J_s =2$ and $J \to 0$  as well as $J =2$ and $J_s \to \infty$:
\begin{equation}
\ket{\Phi_2} \to -\frac{1}{\sqrt{2}} \Bigl( \ket{0101}-\ket{1010} \Bigr).
\end{equation}
From this we can conclude that there must be besides the 2-qubit concurrences
an additional 3- and/or 4-qubit entanglement.\\ 
For the concurrences one finds:
\begin{multline}
C_{12}=C_{14}=C_{23}=C_{34}=\\
\text{max}\Bigl\{0,\frac{2J}{\delta}-\frac{4J^2}
{8\delta^2+(2J_s-J+\delta)^2} \Bigr\}
\end{multline}
$C_{12}, C_{14}, C_{23}$ and $C_{34}$ are greater or equal than 0 for
$J_s=2, J>0$ resp. $J=2, J_s >0$. The exact analytic representation
of $C_{13}$ and $C_{24}$ is only possible in the parameter ranges
$J=2, J_s<2$ and $J_s=2, J>2$:
\begin{multline}
C_{13}=C_{24}=\\
\text{max}\Bigl\{0,\frac{1}{\sqrt{2}\delta}\bigl(\sqrt{\delta^2+\delta(J-2J_s)-4J^2}-\\
\sqrt{\delta^2-\delta(J-2J_s)-4J^2} \bigr)\Bigr\}
\end{multline}
For $J=2, J_s \geq 1$ and $J_s=2, J \leq 4$, $C_{13}$ and $C_{24}$ are equal to 0.\\
In fig. \ref{phi15-erg} we have plotted the sum of the squares
of the concurrences. We can see that the maximum is at $J_s=J=2$. 
It drops for very large $J_s$ to 0 while for $J$ to infinity it levels to a finite value.
Furthermore, it is found that the total global entanglement is constantly 1 
\begin{equation}
Q(\ket{\Phi_2})=1,
\end{equation} 
which means that there is no differentiation between the different qubit entanglements 
in this measure. Again it is very remarkable that the Mermin-Klyshko optimized
operator is describing the additional entanglement (besides the 2-qubit concurrences)
and follows parallel to the curve $1-\sum C_{ij}^2$.
We therefore conclude in analogy to the 3-qubit case that $F_4$ measures the true
3- and 4-qubit entanglements for this state.
This is the most interesting result of our paper, since here is a
quantitative measure of $n$-qubit entanglement $(n = 3,4)$ for a 4-qubit state,
although we cannot discriminate between 3- and 4-qubit entanglement.\\
Again if we look for the structure of $F_4$ we find that all polynomial 
contributions for the limits $J_s=2, J=0$ and $J=2, J_s \to \infty$ come from products of $\sigma^x$
and $\sigma^y$ for the different qubits  
and that the weight of this contribution changes with the strength of 
the additional entanglement (cp. Table \ref{bellparameter}).\\
In a future paper the polynomial structure is investigated in detail.

\section{Conclusions and discussions}

It is shown in this paper that the optimized Mermin-Klyshko operators can be
used very effectively to describe the degree of entanglement in different clusters
of Heisenberg spins. In those cases where there is in the 3-qubit system besides
the concurrences no additional entanglement (i. e. the tangle $\tau =0$) the optimized
$F_3$ operator perfectly describes the 2-qubit entanglement of the system as a
function of the anisotropic parameter in the Heisenberg cluster and it is more or 
less identical to the global entanglement measure 
resp. the sum of the squared concurrences, cp. (\ref{qconctau}).
In those cases where in addition to the 2-qubit concurrences there is a finite
tangle $\tau$, we find that this additional 3-qubit entanglement measured by $\tau$
is nearly perfect described by the optimized $F_3$, as shown in fig. \ref{erg-ps78}.\\
We therefore test this result in a 4-qubit system and again we find two
different cases. We discuss a special state $\ket{\Phi_1}$ where the sum
of the 2-qubit concurrences is mainly proportional to the global entanglement measure
and from this we find that the optimized $F_4$ operator follows this $Q$ value.
In the second eigenstate for the system  $\ket{\Phi_2}$ where the
global entanglement is just equal to 1, independent of the parameters, we expect
besides the 2-qubit concurrences an additional entanglement and this seems to be perfectly 
the case, especially when the expression $1- \sum C_{ij}^2$ is compared in its parameter
dependence to the optimized $F_4$, shown in fig. \ref{phi15-erg}. Our special interest
is here that even at the minima of these functions, at the point $J=J_s=2$, there is
a small but finite higher entanglement which of course at the moment could not be 
interpreted as 3- or otherwise 4-qubit entanglement.\\
It should be noted that the optimization procedure for the $F_N$ operators heavily depends on
the starting values and therefore a procedure has to be used where a random choice for the
starting values has to be done. Another remark is, that this optimization yields much more than
one minimum or maximum and therefore one should be careful with the interpretation of these 
parameters. But at least for the GHZ-state with 4-qubits we have shown that these operators
contain besides the usually used $\sigma^y$ operators additional $\sigma^x$ contributions
\cite{Jaeger:03}. \\
Further work is in preparation where a more extensive study of these $F_N$ operators will be 
presented.


\end{document}